\documentstyle[sprocl,epsfig]{article}

\def\Journal#1#2#3#4{{#1} {\bf #2}, #3 (#4)}

\def\NPA{{\em Nucl. Phys.} A}

\def\PRL{\em Phys. Rev. Lett.}
\def\PRC{{\em Phys. Rev.} C}

\def\be{\begin{equation}}
\def\ee{\end{equation}}
\def\bea{\begin{eqnarray}}
\def\eea{\end{eqnarray}}

\begin{document}
   
\title{ISOVECTOR MULTIPHONON EXCITATIONS IN NEAR SPHERICAL NUCLEI}

\author{N. A. SMIRNOVA}
\address{Centre de Spectrom{\'e}trie Nucl{\'e}aire et de Spectrom{\'e}trie 
de Masse, \\
IN2P3-CNRS and Universit{\'e} Paris Sud, 
B{\^a}t. 104, F-91405 Orsay Campus, France \\
Instituut voor Kern- en Stralingsfysica, University of Leuven, \\ 
Celestijnenlaan 200 D, 3001 Leuven, Belgium\\
E-mail: Nadya.Smirnova@fys.kuleuven.ac.be}
\author{N. PIETRALLA}
\address{Institut f\"ur Kernphysik, Universit\"at zu K\"oln, 
50937 K\"oln, Germany \\  
Wright Nuclear Structure Laboratory, Yale University, 
New Haven, 06520 CT} 
\author{T. MIZUSAKI}
\address{Department of Law, Senshu University, 
1-1, Higashimita 2-chome, Tama-ku, Kawasaki-shi, 
Kanagawa, 214-8580, Japan}
\author{P. VAN ISACKER}
\address{GANIL, BP 5027, F-14076, Caen Cedex 5, France } 

\maketitle\abstracts{ 
The lowest isoscalar and isovector quadrupole and octupole excitations
in near spherical nuclei are studied within the 
the proton--neutron version of the interacting boson model 
including quadrupole and octupole bosons ($sdf$--IBM-2).
The main decay modes of these states in near spherical nuclei are discussed.
}

\section{Introduction} 

The low-lying collective $J^\pi = 2^+$ and $3^-$ excitations in 
near spherical nuclei can be considered as quadrupole and 
octupole vibrations, which represent the most important vibrational 
degrees of freedom at low energies. 
The bosonic phonon concept suggests the occurrence of multiphonon 
states at  excitation energies of $n$ times the one-phonon energy. 
A two-quadrupole-phonon ($2^+ \otimes 2^+$) triplet of states 
($J^\pi = 0^+, 2^+, 4^+$) is usually known in near spherical nuclei. 
Multi-quadrupole-phonon states, 
($3^-\otimes 3^-$) double-octupole states, and double giant dipole 
resonances are actively debated in the literature.

Another interesting example of two-phonon excitations is
the ($2^+ \otimes 3^-$) quadrupole-octupole coupled quintuplet 
($J^\pi = 1^-$ -- $5^-$).
The two-phonon $1^-$ state has been well investigated in  
magic and near spherical nuclei, 
while there are only few examples, where besides the $1^-$ state 
other multiplet members have been identified experimentally 
(see e.g. \cite{SmPi00} for references).
The ($2^+ \otimes 3^-$)-states decay 
besides collective $E2$ and $E3$ transitions to 
the $3^-_1$ octupole phonon state and to the $2^+_1$ 
quadrupole phonon state, respectively, by 
relatively strong $E1$ transitions (see the schematic decay 
pattern in Fig.~1a). This fact was considered \cite{NPE199}
as the evidence of quadrupole-octupole collectivity.

All the low-energy one- and two-phonon states mentioned above 
involve isoscalar phonons. 
Another class of low-lying collective states,
namely the isovector quadrupole excitations in the valence shell, 
or mixed-symmetry states~\cite{IaAr87},  
has been identified in some nuclei \cite{NPCame}. 
The fundamental $2^+_{\rm ms}$ state decays by a weakly collective $E2$ 
transition to the ground state and by a strong $M1$ transition 
to the isoscalar $2^+_1$ state ($B({\rm M1})\sim 1 \mu_N^2$) \cite{IaAr87}. 
In an harmonic phonon coupling scheme one can expect also the existence 
of mixed-symmetry two-phonon multiplets, that involve at least one 
excitation of the mixed-symmetry quadrupole phonon (Fig.~1b). 
Two members of the ($2^+_1\otimes 2^+_{\rm ms}$) quintuplet, 
$J^\pi = 1^+$ and $3^+$ states, are already identified  \cite{PiFr} 
in the near spherical nucleus $^{94}$Mo.
\begin{figure}[t]
\epsfig{figure=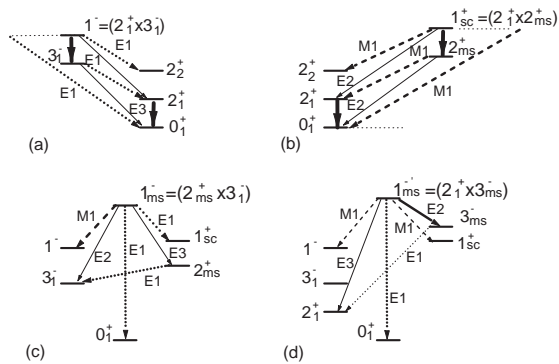,height=2.2in}
\caption{Expected decays of fundamental one-phonon and 
         two-phonon states in the \protect$sdf$-IBM-2. 
         \protect$2^+_{\rm ms} \to 3^-$ transitions were observed 
         already experimentally \protect\cite{Van95,Franpc}.}
\end{figure}
In this contribution we discuss the variety of possible proton-neutron 
non-symmetric excitations in the presence of quadrupole and octupole 
degrees of freedom 
and their decay properties as predicted by the $sdf$--IBM-2 \cite{SmPi00}.

\section{IBM description of isovector excitations}

The $sdf$--IBM-2 \cite{SmPi00} is the version of the IBM ~\cite{IaAr87},  
which considers $s$ ($l^\pi = 0^+$), $d$ ($l^\pi = 2^+$) and 
$f$ ($l^\pi = 3^-$) proton and neutron bosons.
A realistic Hamiltonian relevant for the description of 
near spherical nuclei can have the form
\begin{equation}
\label{H}
\hat{H}= \epsilon_d \hat{n}_d + \epsilon_f \hat{n}_f 
       + \alpha \hat{P}^+\hat{P} + \beta \hat L^2+ \lambda \hat{M}_{13}
       + \lambda' \hat{M}_6 + \lambda ''\hat{M}_7,    
\end{equation}  
where the first four terms provide the spectrum of isoscalar quadrupole and
octupole excitations \cite{IaAr87}, 
while the last three terms called Majorana operators
determine the excitation energies of proton-neutron 
non-symmetric states \cite{IaAr87,SmPi00}.
In case of no octupole bosons, the model reduces 
to the standard IBM-2 \cite{IaAr87}.

The states of Eq.~(\ref{H}) can be classified
by the following group reduction chain quantum numbers
(see \cite{SmPi00} for details). 
\begin{equation}
\label{DS}
\begin{array}{ccccccccc}
{\mbox{U}_{\pi }(13)}&{\otimes }&{\mbox{U}_{\nu }(13)}&{\supset }&
{\mbox{U}_{\pi \nu }(13)}&{\supset }&
{\mbox{U}_{\pi \nu }(6)}&{\otimes }&{\mbox{U}_{\pi \nu }(7)}  . \\ 
\downarrow & & \downarrow & & \downarrow & & \downarrow & & \downarrow
 \\[0pt]     
{[N_{\pi }]} & & {[N_{\nu }]} & & {[N_1,N_2]} & &
{[n_1,n_2]} & & {[m_1,m_2]}  \\     
\end{array}
\end{equation}    

The isoscalar excitations (proton--neutron symmetric ones) 
are described by the totally symmetric 
irreducible representations of the groups from Eq.~(\ref{DS}). 
For these states the $sdf$-IBM-2 reduces to the $sdf$-IBM-1 \cite{IaAr87}.

The two-rowed representations of the U$_{\pi \nu}$(13) group
$[N-1,1]$ correspond to the simplest proton-neutron 
mixed-symmetry states. 
From the reduction 
U$_{\pi \nu }$(13)$\supset $U$_{\pi \nu }$(6)$\otimes $U$_{\pi \nu }$(7), 
we can distinguish three main types of these states:\\
\noindent
1. The $[N-1,1] \supset [N-m-1,1] \otimes [m]$ 
decomposition corresponds to the usual
mixed-symmetry states in the $sd$-space, coupled to $m$ octupole-bosons 
with a wave function,  which is symmetric in the $f$-sector. 
Examples are all the states provided by the IBM-2 \cite{IaAr87}, 
such as the fundamental $2^+_{\rm ms}$ state and the $1^+_{\rm sc}$ 
scissors mode ($m=0$).
The coupling to the octupole bosons give rise to new 
mixed-symmetry states with negative parity. 
We denote the lowest-lying mixed-symmetry two-phonon quintuplet with negative 
parity as $(2^+_{\rm ms} \otimes 3^-_1)$. 
These states are generated by the 
coupling of the lowest $2^+_{\rm ms}$ state in the $sd$-sector and one 
symmetric $f$-boson ($m=1$), and they should 
show a very characteristic decay pattern.
For instance, the $J^\pi = 1^-_{\rm ms}$ state  (see Fig.~1c)
should decay by relatively strong $E1$, strong $M1$, weakly-collective $E2$, 
and collective $E3$ transitions to the ground and 
to the $1^+_{\rm sc }$ states,
to the isoscalar $1^-$ two-phonon state, and to the 
one-phonon $3^-$ and $2^+_{\rm ms}$ states, respectively.\\
\noindent 
2. The $[N-1,1] \supset [N-m]\otimes [m-1,1]$ reduction describes 
the mixed-symmetry states in the $f$-sector and are not considered here. \\
\noindent  
3. Finally, $[N-1,1] \supset [n]\otimes [m]$ states 
are symmetric separately in the $sd$- and in the $f$-sectors, 
but coupled in a non-symmetric way within the full $sdf$-space. 
The simplest example is the $3^-_{\rm ms}$ state 
($N-1$ $s$-bosons and $m=1$ $f$-boson),
which is the mixed-symmetry analogue of the symmetric $3^-_1$  state. 
It should decay by a weakly collective $E3$ transition to the ground state
and by a strong $M1$ transition to the $3^-_1$ state.
Higher excited mixed-symmetry states of this type can be obtained by 
replacing some of the $s$-bosons by $d$-bosons along with appropriate 
angular momentum coupling. 
We can schematically denote the lowest quintuplet of one $d$-boson states 
as ($2^+_1 \otimes 3^-_{\rm ms}$).
The characteristic decay properties of the lowest $1^{- \prime}_{\rm ms}$ 
member of this class are shown in Fig.~1d.

Electromagnetic decay rates of the newly 
predicted mixed-symmetry states with negative parity can be 
derived analytically \cite{SmPi00} in the dynamical symmetry limits of the 
$sdf$-IBM-2. 
For instance the ratio of the quadrupole-octupole collective $E1$ 
excitation strengths to the $1^-_{\rm ms}$ state and to the well 
known symmetric two-phonon $1^-_1$ state is predicted in the U(5) limit 
to be 
\begin{displaymath}
\frac{B(E1;0^+_1 \rightarrow 1^-_{\rm ms})}
     {B(E1;0^+_1 \rightarrow 1^-_1)}  = 
     \frac{N(N-2)}{(N-1)^2}\ 
     \frac{(1-\eta^2)N_\pi N_\nu}{(N_\pi + \eta N_\nu)^2} \ , 
\end{displaymath} 
where $\eta \equiv e_\nu/e_\pi$ is the ratio 
of the $E2$ effective boson charges  \cite{SmPi00}. 

\section{Application to experiment}

The non-symmetric one- and two-phonon quadrupole excitations
have been observed and studied in a number of near spherical nuclei. 
Their coupling to the octupole degree of freedom is an unavoidable 
prediction of the $sdf$-IBM-2. 
The perhaps most easily observable mixed-symmetry states with negative 
parity should be the $(2^+_{\rm ms} \otimes 3^-)$ states.
Both constituting phonons (isovector quadrupole and isoscalar octupole)
are known in some nuclei, i.e. $^{142}$Ce,  $^{144}$Nd, $^{94}$Mo,
 $^{134,136}$Ba. 
The excitation energy of the new states is expected to be close 
to the sum energy of the constituting phonons, namely around 3.5--4 MeV. 

\section*{Acknowledgments}
We thank P. von Brentano and N.V. Zamfir for discussions. 
This work was supported by the DFG under Contract no. Pi 393/1-1 
and partially by the U.S. DOE under Contract no. 
DE-FG02-91ER-40609.

\end{document}